\documentclass[a4paper,fleqn,usenatbib]{mnras}

\usepackage[T1]{fontenc}
\usepackage{ae,aecompl}

\usepackage{graphicx}	
\usepackage{amsmath}	
\usepackage{amssymb}	
\usepackage{hyperref}    

\title[Multiple populations in three SMC clusters]{The Search for Multiple Populations in Magellanic Cloud Clusters II: The Detection of Multiple Populations in Three Intermediate-Age SMC Clusters\thanks{Based on observations made with the NASA/ESA Hubble Space Telescope, and obtained from the Hubble Legacy Archive, which is a collaboration between the Space Telescope Science Institute (STScI/NASA), the Space Telescope European Coordinating Facility (ST-ECF/ESA) and the Canadian Astronomy Data Centre (CADC/NRC/CSA).}}

\author[F. Niederhofer et al.]{
F. Niederhofer$^{1}$\thanks{FN: fniederhofer@stsci.edu}, 
N. Bastian$^{2}$,
V. Kozhurina-Platais$^{1}$,
S. Larsen$^{3}$,
K. Hollyhead$^{2}$,
\newauthor
C. Lardo$^{2}$,
I. Cabrera-Ziri$^{2,4}$,
N. Kacharov$^{5}$,
I. Platais$^{6}$,
M. Salaris$^{2}$,
M. Cordero$^{7}$,
\newauthor
E. Dalessandro$^{8,9}$,
D. Geisler$^{10}$,
M. Hilker$^{4}$,
C. Li$^{11}$,
D. Mackey$^{12}$,
and  A. Mucciarelli$^{8}$,
\\
$^{1}$ Space Telescope Science Institute, 3700 San Martin Drive, Baltimore, MD 21218, USA \\
$^{2}$ Astrophysics Research Institute, Liverpool John Moores University, 146 Brownlow Hill, Liverpool L3 5RF, UK \\
$^{3}$ Department of Astrophysics/IMAPP, Radboud University, P.O. Box 9010, 6500 GL Nijmegen, The Netherlands \\
$^{4}$ European Southern Observatory, Karl-Schwarzschild-Stra\ss e 2, D-85748 Garching bei M\"unchen, Germany \\
$^{5}$ Max-Planck-Institut f\"ur Astronomie, K\"onigstuhl 17, D-69117 Heidelberg, Germany \\
$^{6}$ Department of Physics and Astronomy, Johns Hopkins University, 3400 North Charles Street, Baltimore, MD 21218, USA\\
$^{7}$ Astronomisches Rechen-Institut, Zentrum f\"ur Astronomie der Universit\"at Heidelberg, M\"onchhofstra\ss e 12-14, D-69120 Heidelberg, Germany \\
$^{8}$ Department of Physics and Astronomy, University of Bologna, Viale Berti Pichat 6/2, I-40127 Bologna, Italy \\
$^{9}$ INAF - Osservatorio Astronomico di Bologna, via Ranzani 1, 40127, Bologna, Italy  \\
$^{10}$ Departamento de Astronomia, Universidad de Concepcion, Casilla 160-C, Chile \\
$^{11}$ Department of Physics and Astronomy, Macquarie University, Sydney, NSW 2109, Australia \\
$^{12}$ Research School of Astronomy and Astrophysics, Australian National University, Canberra, ACT 2611, Australia
}

\date{Accepted XXX. Received YYY; in original form ZZZ}

\pubyear{2016}

\begin{document}
\label{firstpage}
\pagerange{\pageref{firstpage}--\pageref{lastpage}}
\maketitle

\begin{abstract}
This is the second paper in our series about the search for multiple populations in Magellanic Cloud star clusters using the Hubble Space Telescope. Here we report the detection of multiple stellar populations in the colour-magnitude diagrams of the intermediate-age clusters Lindsay~1, NGC~416 and NGC~339. With ages between 6.0 and 7.5~Gyr, these clusters are the youngest ones in which chemical abundance spreads have been detected so far. 
This confirms that the appearance of multiple populations is not restricted to only ancient globular clusters, but may also be a common feature in clusters as young as 6 Gyr.
Our results are in agreement with a recent spectroscopic study of Lindsay~1. We found that the fraction of enriched stars in NGC~416 is $\sim$45\% whereas it is $\sim$25\% in NGC~339 and $\sim$36\% in Lindsay~1. Similar to NGC~121, these fractions are lower than the average value for globular clusters in the Milky Way.
\end{abstract}




\section{Introduction}
\label{sec:intro}

The increasing precision in observing techniques has revolutionized our view of star clusters in the last few decades. It is now well established that old globular clusters (GCs) are not, as previously thought, simple stellar populations but are composed of multiple populations. The stars of the various populations show different (anti-)correlated abundances of light elements, the most prominent being the Na-O and C-N anti-correlations \citep[e.g.][]{Carretta09a, Cannon98}. Moreover, some clusters additionally show a Mg-Al anti-correlation \citep[e.g.][]{Carretta09b}. 

Multiple populations seem to be an inherent and universal property of GCs (with the only known possible exceptions being IC~4499 and Ruprecht 106 - 
\citealt{Walker11,Villanova13}) and their appearance seems not to depend on the environment of the cluster and the type of the host galaxy. Besides the extensively studied GCs in the Milky Way \citep[see e.g.][for a review]{Gratton12}, also GCs in nearby dwarf galaxies host multiple populations, for example in the Fornax dwarf spheroidal galaxy \citep{Larsen14}, the Sagittarius dwarf galaxy \citep{Carretta10, Carretta14} and the Large Magellanic Cloud (LMC, \citealt{Mucciarelli09}). Recently, \citet{Dalessandro16} and \citet[][hereafter Paper I]{Niederhofer17} also detected multiple populations in NGC~121, the only 'classical' GC in the Small Magellanic Cloud (SMC). This cluster has an age of about 10.5~Gyr \citep{Glatt08a}.

It is, however, still unclear if the appearance of multiple populations depends on when or how long ago the cluster formed. To date, there have been no clusters younger than 3~Gyr found with chemical abundance variations  \citep[see e.g.][]{Mucciarelli08,Mucciarelli11,Mucciarelli14,Davies09, Cabrera-Ziri16}. Clusters that fill the age gap between 3~Gyr and 10 Gyr have to date been almost completely neglected by studies searching for multiple populations. So far, only \citet{Hollyhead16} have studied a cluster that falls in this range of ages. They spectroscopically analyzed a sample of red giant branch (RGB) stars in the $\sim$7.5~Gyr old \citep{Glatt08b} cluster Lindsay~1 and found a significant variation in N amongst their sample of stars. This result provides the first evidence that multiple populations are likely to be present in clusters younger than about 10~Gyr.

Recently, we started a photometric survey of star clusters in the Magellanic Clouds covering a large range of ages and masses to search for anomalies in their colour-magnitude diagrams (CMDs) due to possible spreads in chemical abundances, mainly in N and C (see Paper~I for details of the survey). Using this sample of targets, our goal is to establish the lower age limit down to which star clusters show evidence for multiple populations. As a first result of this study, we detected a bifurcation in the RGB of the $\sim$10.5~Gyr old SMC cluster NGC~121 (see Paper~I), consistent with the independent study by \citet{Dalessandro16}. Here, we present the analysis of three more SMC clusters, Lindsay~1, NGC~339 and NGC~416, which have ages between $\sim$6.0 and 7.5~Gyr \citep{Glatt08b}. Table~\ref{tab:cluster_list} lists the basic parameters of the three clusters.

This paper is structured as follows: We briefly describe the observations and data reduction procedures in \S\ref{sec:obs}. The analysis and the results are shown in \S\ref{sec:analysis}. In \S\ref{sec:conclusions} we present the discussion of our results and draw final conclusions.

\begin{table*} 
\centering
\caption{Properties of the three clusters analyzed in this work.  \label{tab:cluster_list}}
\begin{tabular}{l c c c c c c c c} 

\hline\hline
\noalign{\smallskip}
Cluster Name  &  RA & Dec &  Age & Ref. & Mass & Ref. & Metallicity Z & Ref. 
\\
 & & & [Gyr] && [10$^5$~M$_{\sun}$] & & &
\\
\noalign{\smallskip}
\hline
\noalign{\smallskip}
Lindsay~1 & 00$^h$ 03$^m$ 54$^s$.0 & -73$\degr$ 28$\arcmin$ 18$\arcsec$ & 7.5 & (1) & $\sim$2.0 & (2) & 0.001$^{a}$ & (1)
\\
NGC~339 & 00$^h$ 57$^m$ 48$^s$.90 & -74$\degr$ 28$\arcmin$ 00.2$\arcsec$ & 6.0 & (1) & 0.8 & (3) & 0.001$^{a}$ & (1)
\\
NGC~416 & 01$^h$ 07$^m$ 54$^s$.98 & -72$\degr$ 20$\arcmin$ 50.6$\arcsec$ & 6.0 & (1) & 1.6 & (3) & 0.002$^{a}$ & (1)
\\
\noalign{\smallskip}
\hline
\end{tabular}
\\
(1)~\citet{Glatt08b}; 
(2)~\citet{Glatt11}; 
(3)~\citet{McLaughlin05}
\\
$^{a}$From fitting Padova isochrones \citep{Girardi2000,Girardi08} to CMDs in optical filters
\end{table*}

\section{Observations and Data Reduction}
\label{sec:obs}

The observations for the clusters analyzed in this paper are from our ongoing \textit{Hubble Space Telescope} (HST) survey (GO-14069, PI. N. Bastian). All three targets have been observed with the WFC3/UVIS instrument in the near-UV/optical wide-band filters $F336W$ and $F438W$, as well as in the narrow-band filter $F343N$, with long, intermediate and short exposures. These filters are particularly suitable to separate populations with varying N and C abundances in the CMD. The $F336W$ and $F343N$ filters contain a strong NH absorption band whereas the $F438W$ filter has a CH absorption feature within its pass-band (see \citealt{Larsen15}, Paper~I). Additionally, we also use archival observations of NGC~339 in the $F555W$ and $F814W$ filters from the program GO-10396 (PI: J.Gallagher), taken with the ACS/WFC instrument. 

For the photometry, we used the images that have been corrected for imperfect charge transfer efficiency (CTE) and calibrated through the standard HST pipeline for bias, dark, low-frequency flats and new improved UVIS zero-points \citet{Ryan16}. We derived the stellar photometry using the spatially variable "effective point spread function" (ePSF) method (J. Anderson, private communication). The instrumental magnitudes were then transformed into the VEGAMAG system using the newly derived improved UVIS VEGAMAG zero-points from the WFC3 instrument website.
Finally, the derived stellar positions were corrected  for the WFC3/UVIS geometric distortion \citep{Bellini11}. 

The ACS/WFC observations were processed in a similar manner. For the photometry we used the ePSF libraries for ACS/WFC \citep{AndersonKing06}. The transformations of the instrumental magnitudes into the VEGAMAG system were performed with the corrections described in \citet{Sirianni05}.  
We refer the interested reader to Paper~I where the survey and the photometric techniques are described in more detail.

\begin{figure}
\centering
  \includegraphics[width=8cm]{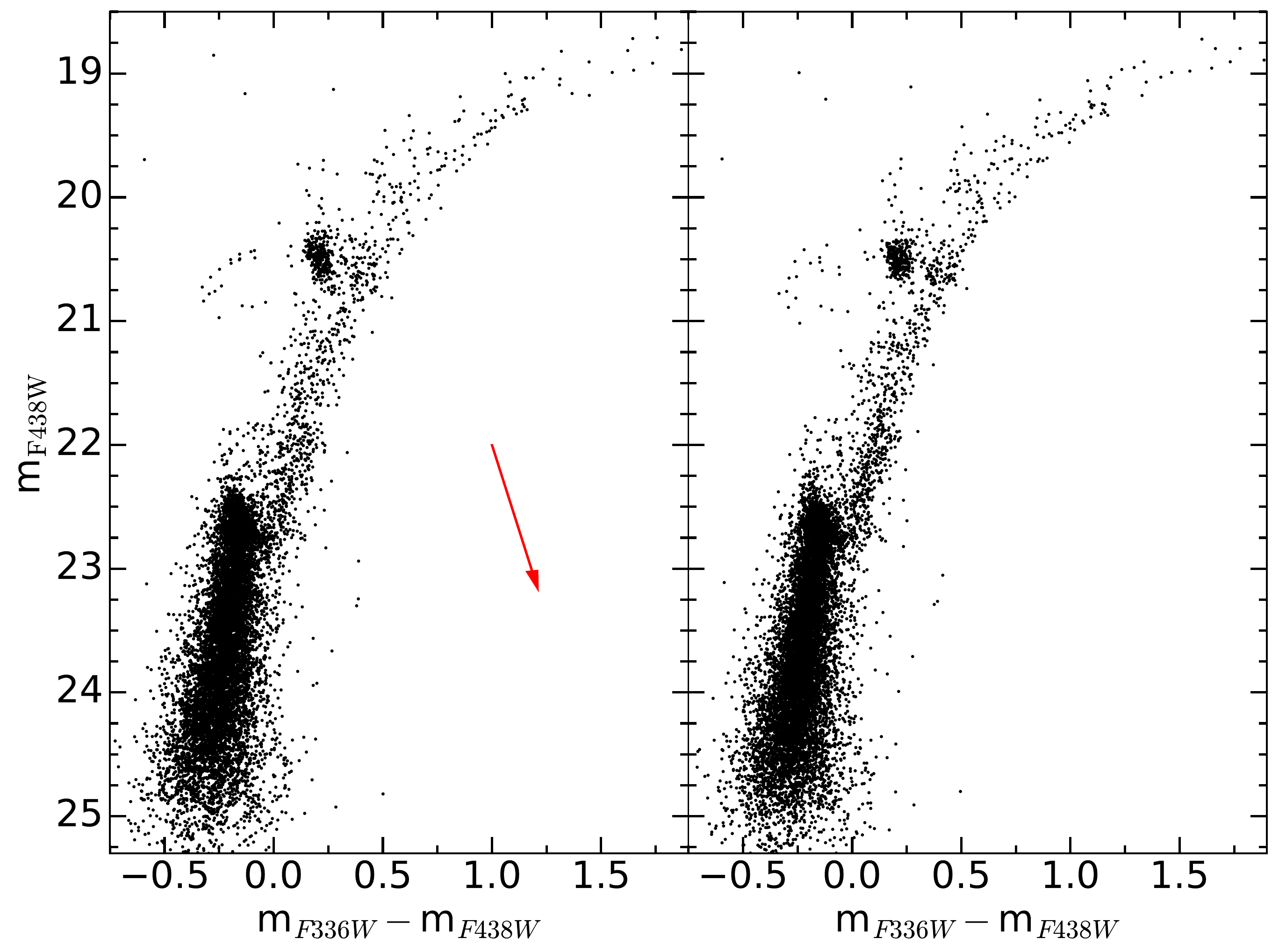}
  \caption{CMD of NGC~416 before (left) and after (right) the differential reddening correction. The direction of the reddening vector in the $m_{F438W}$ vs $m_{F336W}-m_{F438W}$ colour-magnitude space is indicated in the left-hand panel as a red arrow.}
   \label{fig:ngc416}
\end{figure}

Inspecting the CMDs for the clusters, we found that the data for NGC~416 is severely affected by differential reddening across the extent of the cluster. This is reflected in the poorly defined features in the CMD and in the extended red clump morphology along the direction of the reddening vector (see left-hand panel in Figure~\ref{fig:ngc416}). We therefore corrected the data for NGC~416 using the technique that is explained in detail in \citet{Milone12}. The resulting corrected CMD is displayed in the right-hand panel of Figure~\ref{fig:ngc416}. We see that the various features now show less scatter and also the red clump has a more compact shape. We use the corrected magnitudes and colours for our further analysis.

Finally, we statistically subtracted any field star contamination from the cluster CMDs. For this, we selected a field centred on the cluster and a reference field of equal area, maximizing the spatial separation between the two fields. We then removed for each star in the CMD of the reference field the star that is closest in colour-magnitude space (within certain thresholds) from the cluster's CMD \citep[see e.g.][]{Niederhofer15}. This procedure, however, only works for NGC~339 and NGC~416 as they have a compact structure. We could not clean the CMD of Lindsay~1 for field stars as it is too extended to define a reasonable reference field for the subtraction within our single pointing (162$\times$162$\arcsec$ field-of-view; \citealt{Glatt09} reported a core radius of 61$\farcs$7). The field star contamination in the CMD of Lindsay~1 is expected to be minimal, though, as the cluster is located in the outskirts of the SMC.

\section{Analysis}
\label{sec:analysis}

\begin{figure*}
\centering
 \begin{tabular}{c c c}
  \includegraphics[width=5.6cm]{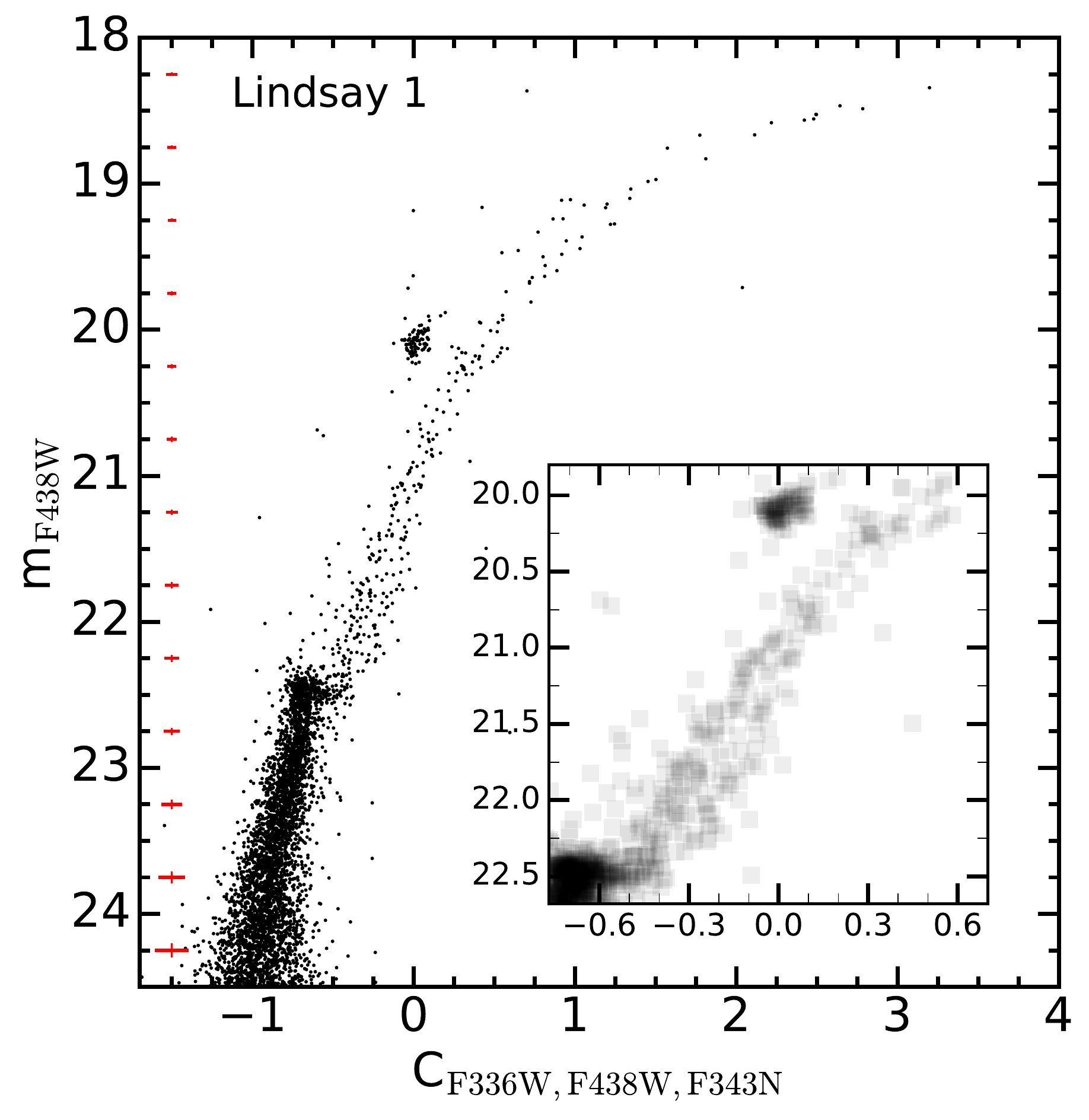} &
  \includegraphics[width=5.6cm]{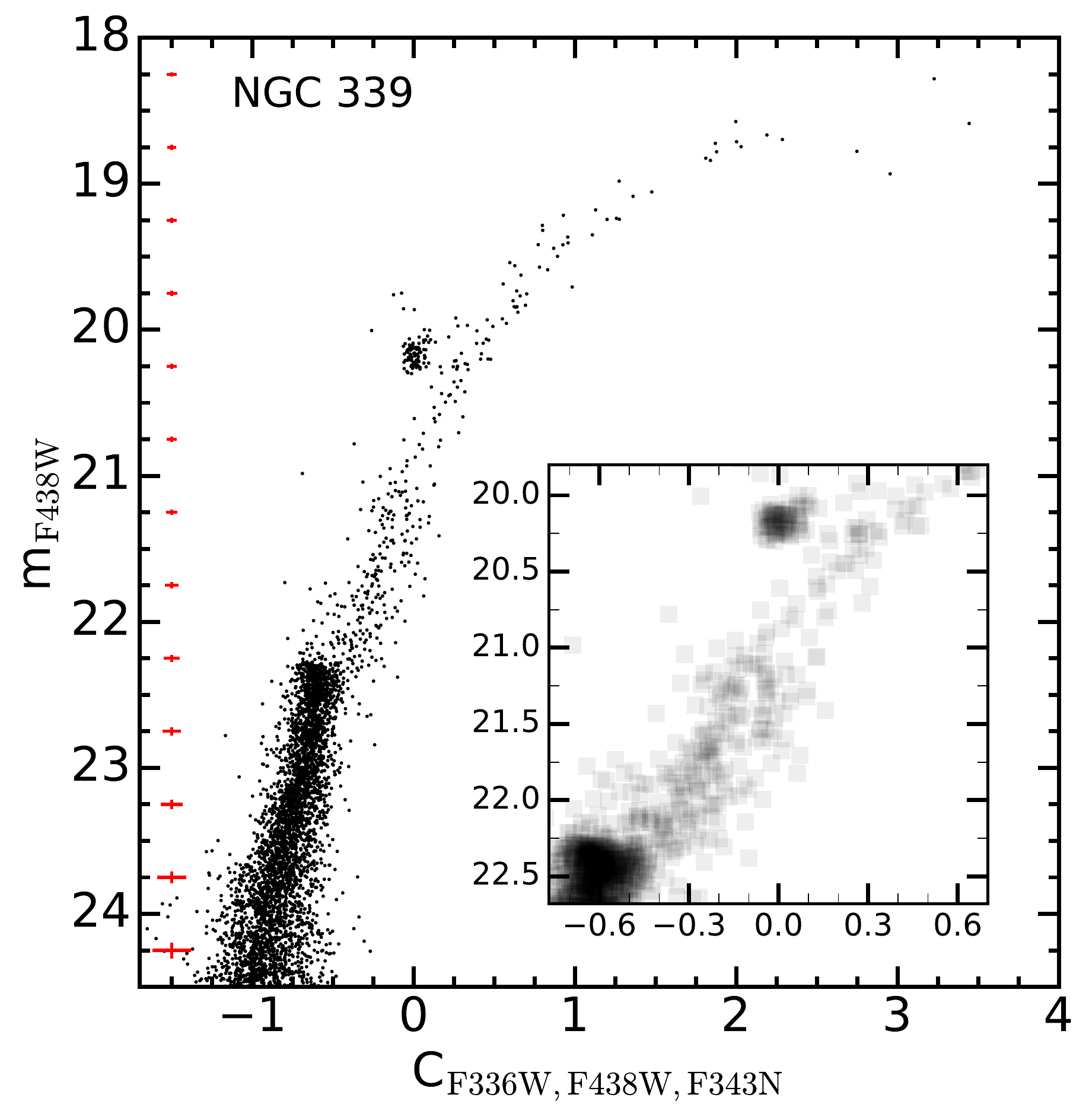} &
  \includegraphics[width=5.6cm]{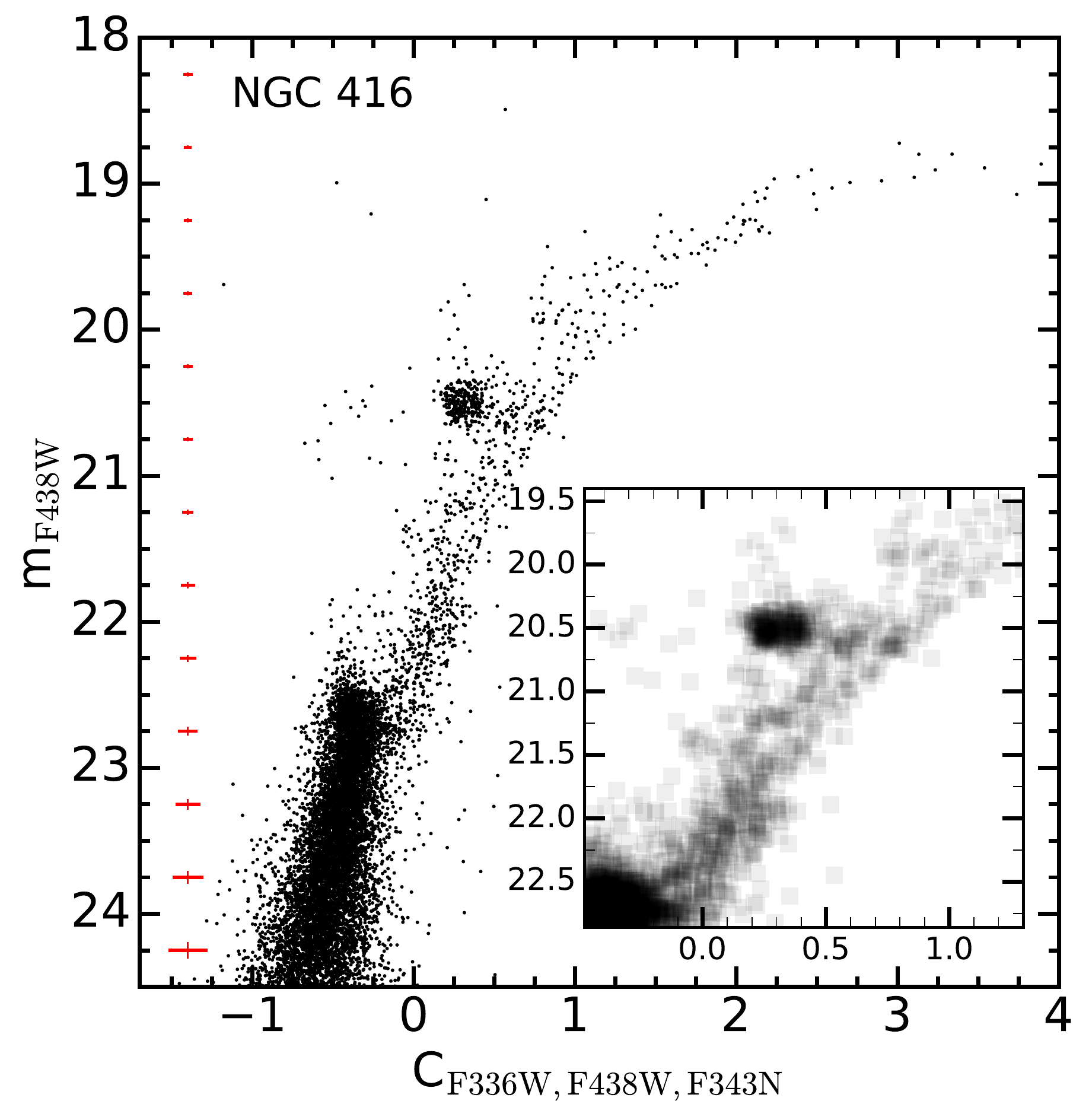} \\
 \end{tabular}
  \caption{$m_{F438W}$ vs $C_{F336W,F438W,F343N}$ CMDs of our sample of clusters, Lindsay~1 (left), NGC~339 (middle) and NGC~416 (right). The inlay in each panel shows a Hess diagram zooming into the region of the RGB where the splitting is most evident. The red crosses show the typical errors in $C_{F336W,F438W,F343N}$ and magnitude as a function of the $F438W$ magnitude. At $m_{F438W}$ = 19, the errors are $\sim$0.004~mag in $F438W$ and $\sim$0.017~mag in $C_{F336W,F438W,F343N}$  and increase to $\sim$0.037~mag in $F438W$ and $\sim$0.090~mag in $C_{F336W,F438W,F343N}$ at $m_{F438W}$ = 24.}
   \label{fig:cmds}
\end{figure*}

In Paper~I, we found that a filter combination of the form $(F336W-F438W)-(F438W-F343N)=C_{F336W,F438W,F343N}$ as the CMD's colour-axis is best suited to separate populations with different C and N abundances. The models predict a splitting in the RGB with this filter combination. Stars with a primordial composition (first population) follow a bluer/brighter sequence whereas stars enriched in N and depleted in C (second population) lie on a redder/fainter track.

\subsection{The Structure of the RGB in Lindsay~1, NGC~399 and NGC~416}

Figure~\ref{fig:cmds} shows the CMDs for the three clusters using this pseudo colour as the abscissa. All clusters reveal a broadening/split in the lower parts of their RGB, indicative of the presence of multiple populations. The inlays in each panel of Figure~\ref{fig:cmds} show a Hess diagram zooming in on the RGB region of the respective cluster. In these diagrams, the splitting of the RGB is most obvious in Lindsay~1 and NGC~416. The CMD of NGC~416 which has the most populated RGB, even shows the presence of two individual RGB bumps at $m_{F438W}\sim$20.6~mag right next to the red clump (Figure~\ref{fig:cmds} right-hand panel). The RGB of NGC~339 is more scattered and does not show a bifurcation, but it is clearly broadened in its lower parts (Figure~\ref{fig:cmds} middle panel). 

\begin{figure}
\centering
 \begin{tabular}{c}
  \includegraphics[width=6.5cm]{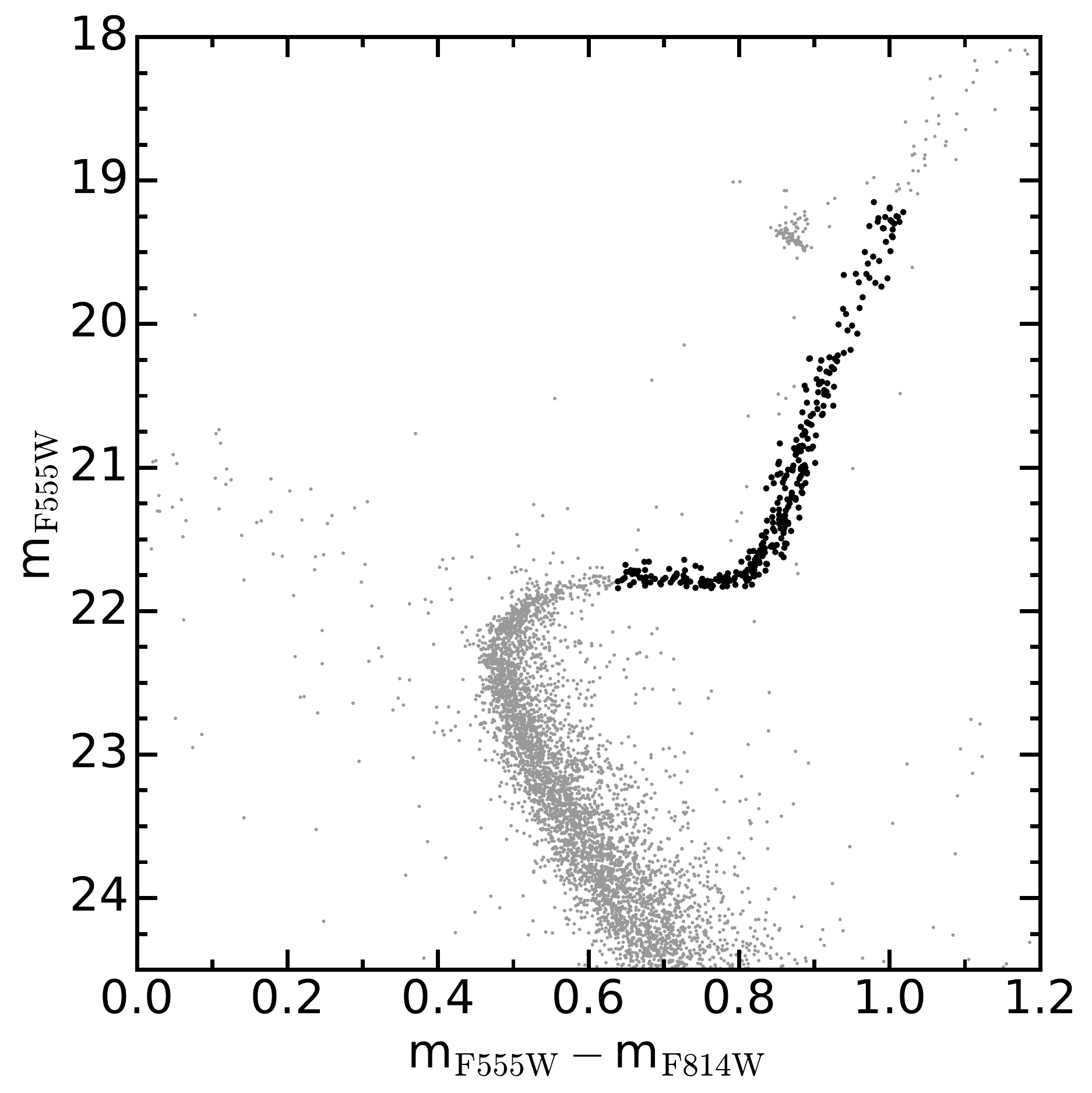} \\
  \includegraphics[width=6.5cm]{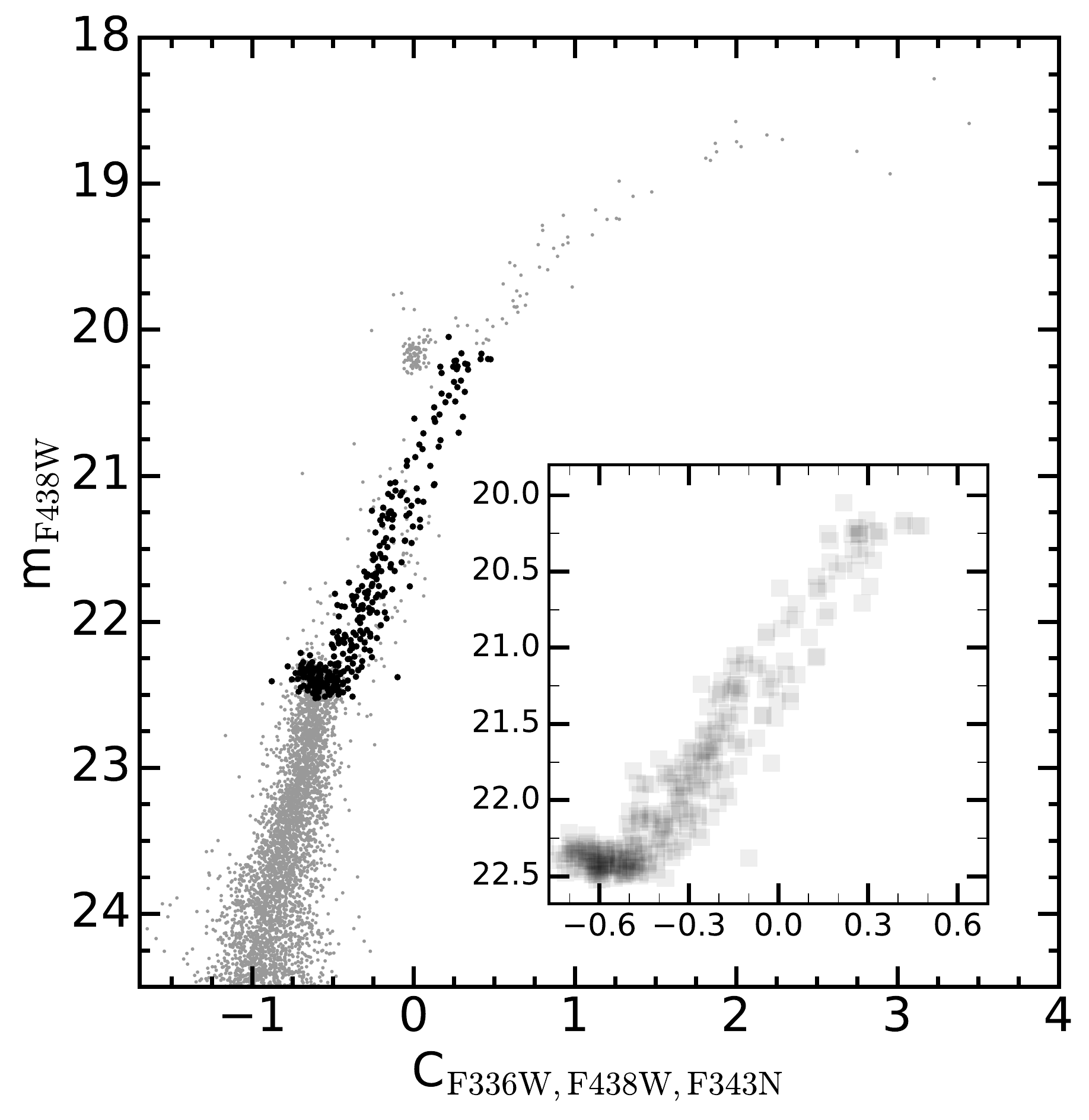} \\
 \end{tabular}
  \caption{Upper panel: $m_{F555W}$ vs $m_{F555W}-m_{F814W}$ CMD of NGC~339. The fiducial RGB stars that we selected for the further analysis are marked as black dots. Lower Panel: $m_{F438W}$ vs $C_{F336W,F438W,F343N}$ CMD of NGC~339. The same stars that we selected in the optical CMD are also indicated here as black dots. The inlay Hess diagram shows only the fiducial RGB stars.}
   \label{fig:ngc339VI}
\end{figure}

The RGB in NGC~339 might still be contaminated by a younger, more metal-rich field star population that was not entirely subtracted by our statistical de-contamination procedure. Such a population crosses the RGB in the used filter combination thus mimicking enriched cluster stars on the lower RGB. The bend-like structure in the RGB at a $F438W$ magnitude of $\sim$21.25 might indicate that the broadened RGB of NGC~339 still contains some contribution of field stars that cannot be discriminated from actual second population stars. This degeneracy can be broken, when using observations in optical filters. In order to get a cleaner sample of stars belonging to the red RGB stars in the $m_{F438W}$ vs $C_{F336W,F438W,F343N}$ CMD of NGC~339 we constructed a CMD using the optical filters $F555W$ and $F814W$ (see upper panel of Figure~\ref{fig:ngc339VI}). In this filter combination, the younger field population lies blue-ward of the main sequence. In a first step, we used the optical filters to statistically subtract the field star contamination form the cluster data. As can be seen in the upper panel of Figure~\ref{fig:ngc339VI}, there is still some contribution from a younger populations left in the CMD. We therefore selected a sample of fiducial RGB stars in the $m_{F555W}$ vs $m_{F555W}-m_{F814W}$ CMD and used this sample for our further analysis. In Figure~\ref{fig:ngc339VI} the selected RGB stars are marked with black dots in both CMDs. The $m_{F438W}$ vs $C_{F336W,F438W,F343N}$ CMD clearly shows that there was indeed a considerable fraction of an unrelated population present in the lower RGB (Figure~\ref{fig:ngc339VI} lower panel). But the lower RGB is still broadened, indicating that there is indeed a second population present in the cluster. 

\begin{figure}
\centering
 \begin{tabular}{c}
  \includegraphics[width=8.5cm]{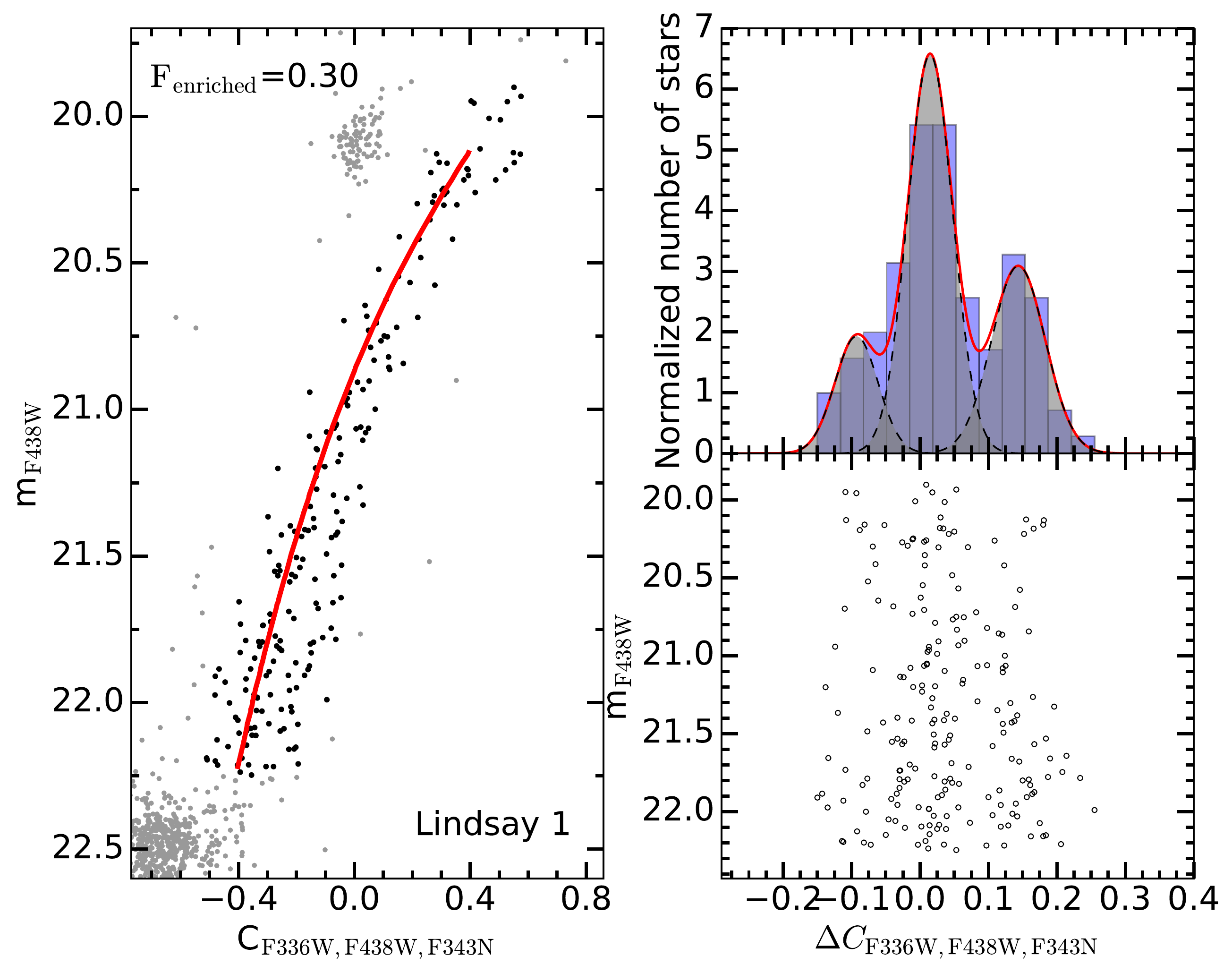} \\
  \includegraphics[width=8.5cm]{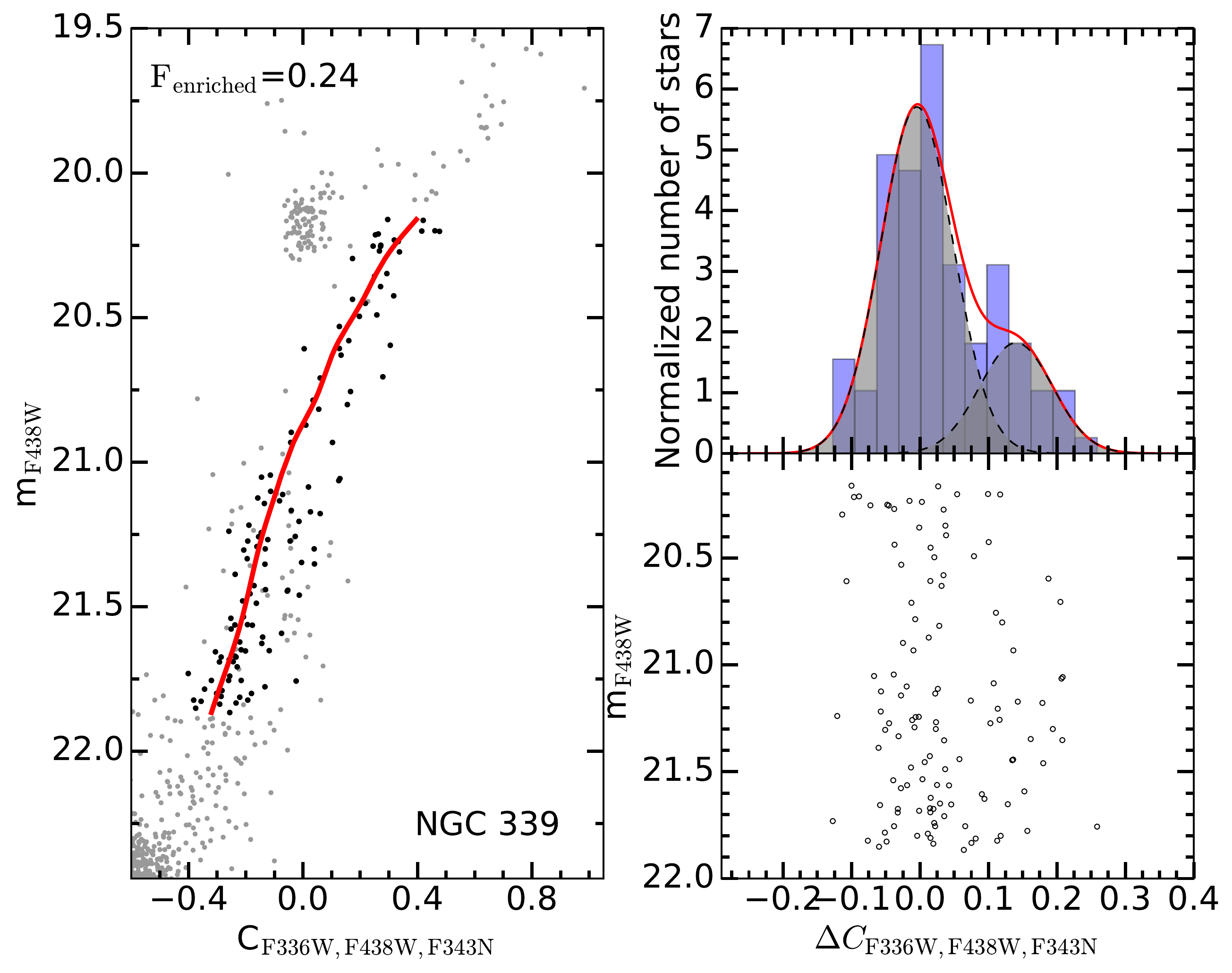} \\
  \includegraphics[width=8.5cm]{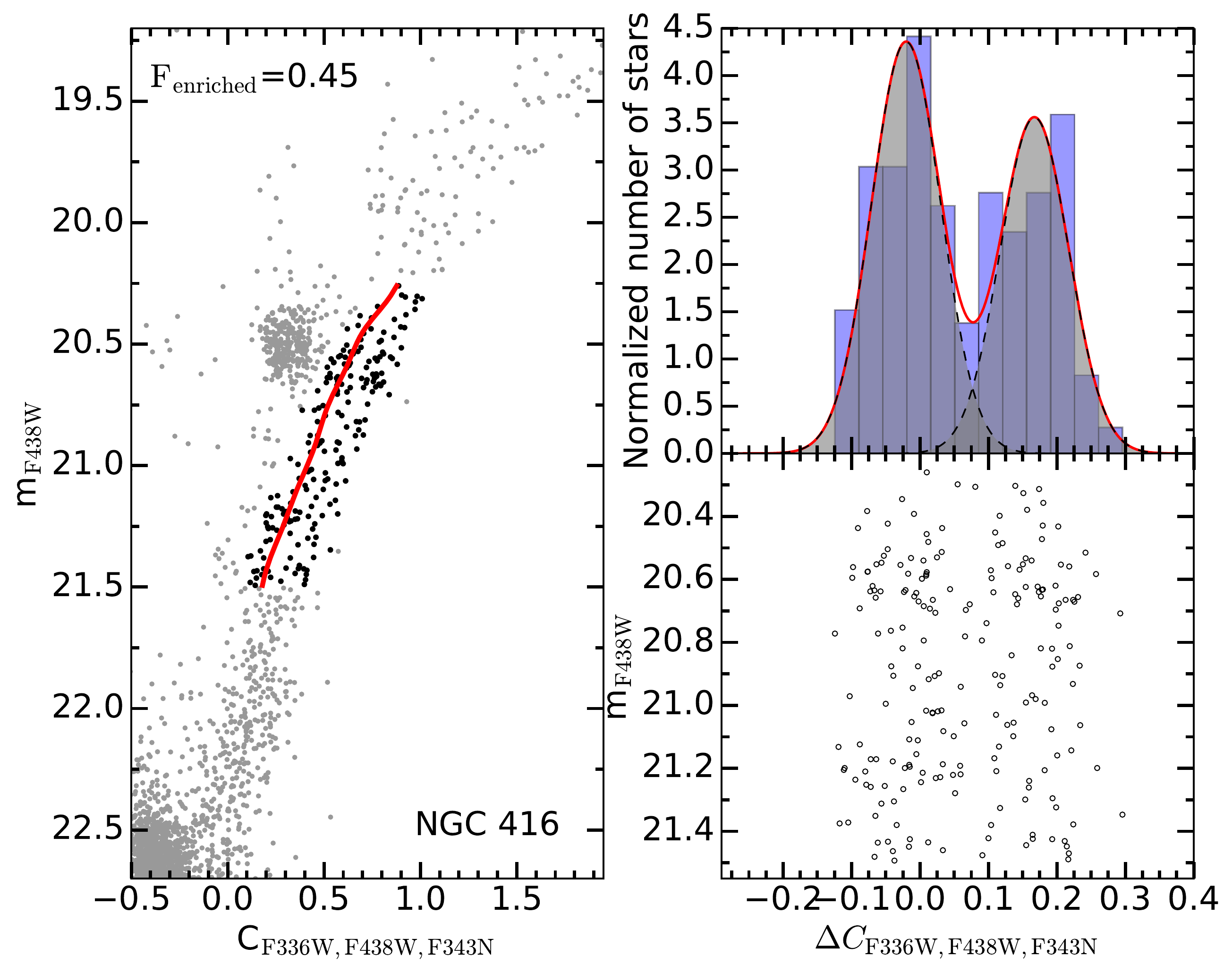} \\
 \end{tabular}
  \caption{Distribution of RGB stars in Lindsay~1 (top panel), NGC~339 (middle panel) and NGC~416 (bottom panel). The left plot in each panel shows a zoom into the RGB region in the $m_{F438W}$ vs $C_{F336W,F438W,F343N}$ CMD. The selected stars for the colour distribution are marked by black filled circles and the red line shows the fiducial line. The lower right plots show the verticalized CMDs of the RGB stars. The abscissa marks the distance of each star, $\Delta C_{F336W,F438W,F343N}$, from the RGB fiducial line. The upper right panels show the resulting histograms of the distributions. We also show here the Gaussian Mixture Models that fit best the unbinned data. The mixture models are shown with a red solid line whereas the individual Gaussians are indicated as grey shaded ares and dashed black lines.}
   \label{fig:rgb}
\end{figure}

In order to analyze the structure of the RGBs in the three clusters we fitted a fiducial line to the blue sequence of the RGB and calculated for each star the distance $\Delta C_{F336W,F438W,F343N}$ from this line (see Figure~\ref{fig:rgb}, the lower right figure of each panel). 
The stars of Lindsay~1 and NGC~416 show a clear bi-modal pseudo-colour distribution. NGC~339 shows a less clear pattern as the RGB stars are more scattered there. But the lower part of the RGB is significantly broadened to the red and this spread is comparable to the separation of the two sequences in NGC~416 and Lindsay~1. 

\subsection{The Fraction of Enriched Stars}

We now use Gaussian Mixture Models (GMMs) with a variable number of components to fit the distribution along the $\Delta C_{F336W,F438W,F343N}$ axis of the individual stars in the three verticalized CMDs. To find the number of Gaussians that reproduce the observed distribution best, we used the Akaike Information Criterion (AIC, \citealt{Akaike74}). The results are shown in the upper right plots of Figure~\ref{fig:rgb}. There, we also show the binned data in the form of a histogram for a comparison. The distribution of the unbinned RGB stars in NGC~339 and NGC~416 is best fitted with a two-component Gaussian whereas the fit to the data of Lindsay~1 favours three components. We define the fraction of enriched stars as the weight of the Gaussian fitting for the red sequence in each cluster. We find that in Lindsay~1 the fraction of enriched stars is 30\%,  in NGC~339 there are 24\% enriched stars and NGC~416 has 45\% stars that belong to the second population. 

The exact fraction of first to second population stars, however, slightly depends on the exact choice of the RGB fiducial line which cannot unambiguously be defined in clusters with a sparsely populated RGB. We note here that the CMD of Lindsay~1 has not been cleaned for any field-star component. There might be a certain fraction of stars not related to the cluster in the part of the CMD that we used for estimating the fraction of enriched stars. However, note that \citet{Parisi16} find that the field stars around Lindsay~1 are generally more metal-poor than the cluster, and would thus presumably lie mostly blue-ward of the cluster RGB, meaning we would overestimate the fraction of primordial Lindsay~1 stars and thus underestimate the fraction of enriched stars. It is also possible that the sparse blue component of the verticalized RGB is due to these metal-poor field stars. When excluding this blue peak from the calculation, we found a fraction of 36\% enriched stars in Lindsay~1.

\begin{figure}
\centering
 \begin{tabular}{c}
  \includegraphics[width=6.5cm]{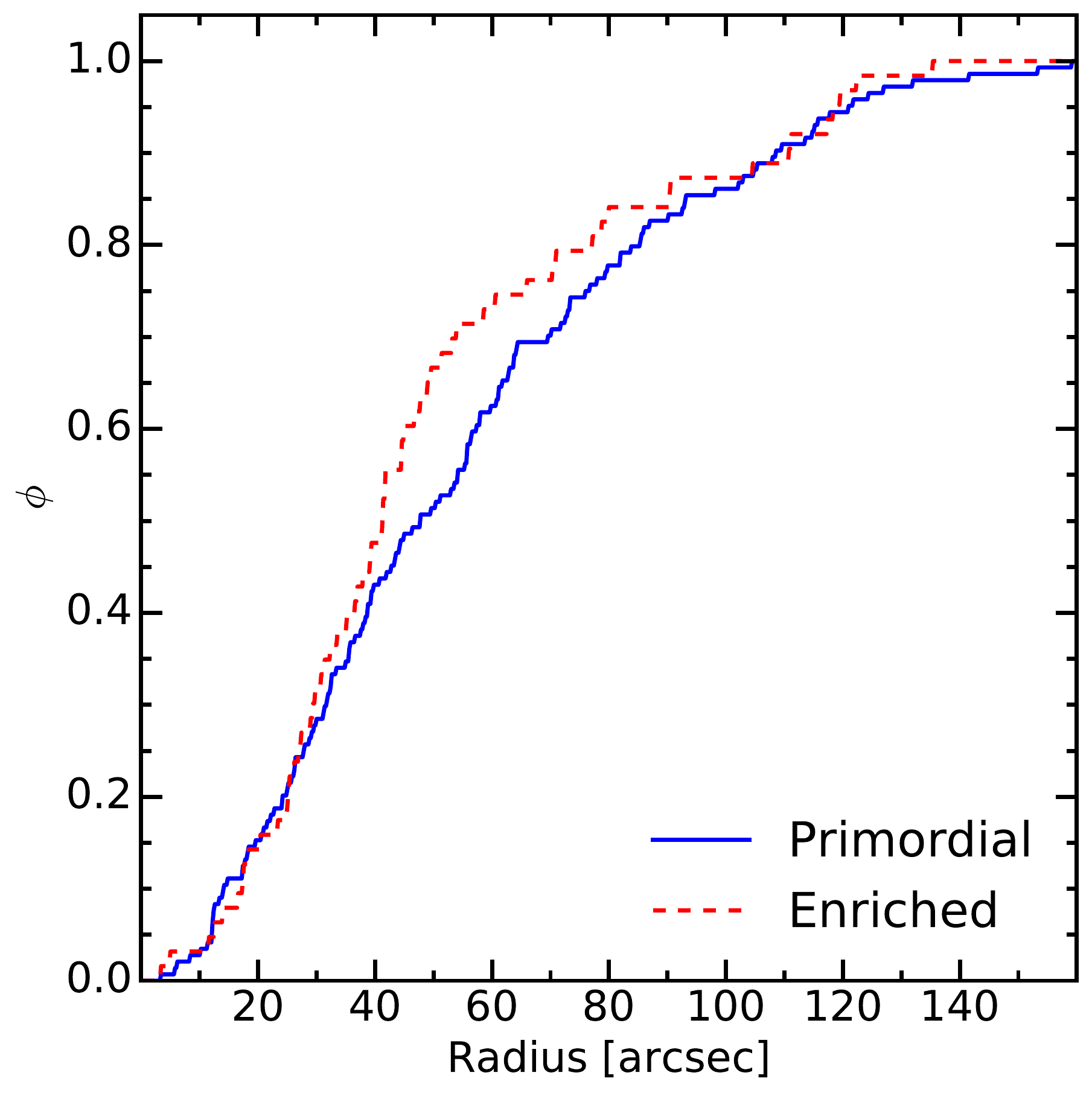} \\
  \includegraphics[width=6.5cm]{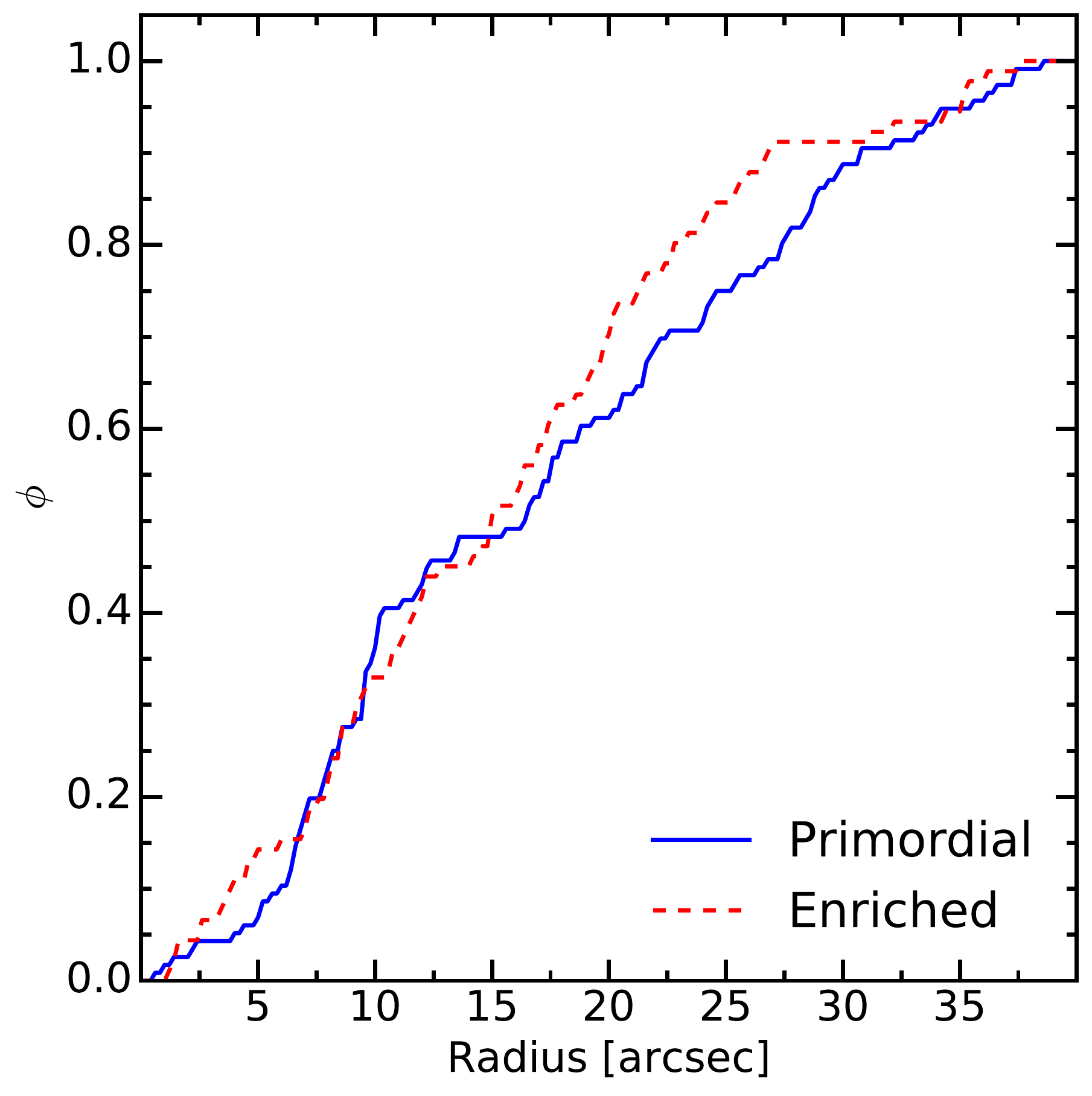} \\
 \end{tabular}
  \caption{Cumulative distributions of the two sequences in Lindsay~1 (top panel) and NGC~416 (bottom panel). The blue solid line indicates the primordial sequence whereas the red dashed line shows the enriched RGB. In both clusters the two branches follow the same trend, indicating that the red sequence is not dominated by a contribution from field stars.}
   \label{fig:cum_dist}
\end{figure}

The fraction of enriched stars in NGC~339 is affected by stars not related to the cluster, as the sequence of the second population contains most likely still a considerable contribution of field stars. Therefore, we used observations in the $F555W$ and $F814W$ filters to get a cleaner sample of RGB stars. We do not have this problem in NGC~416, as we are estimating the fraction of enriched stars above the point where the field star sequence crosses the RGB of the cluster. In order to test if there is a considerable amount of stars within the two RGB branches that do not belong either to Lindsay~1 or NGC~416, we looked at the cumulative distribution of the two sequences as a function of the distance from the cluster centre. These distributions are shown in Figure~\ref{fig:cum_dist}. In both clusters, the red and the blue sequence follow a similar distribution, suggesting that none of the branches is dominated by unrelated field stars.

\subsection{Comparison with Spectroscopic Results}

\begin{figure}
\centering
  \includegraphics[width=6.5cm]{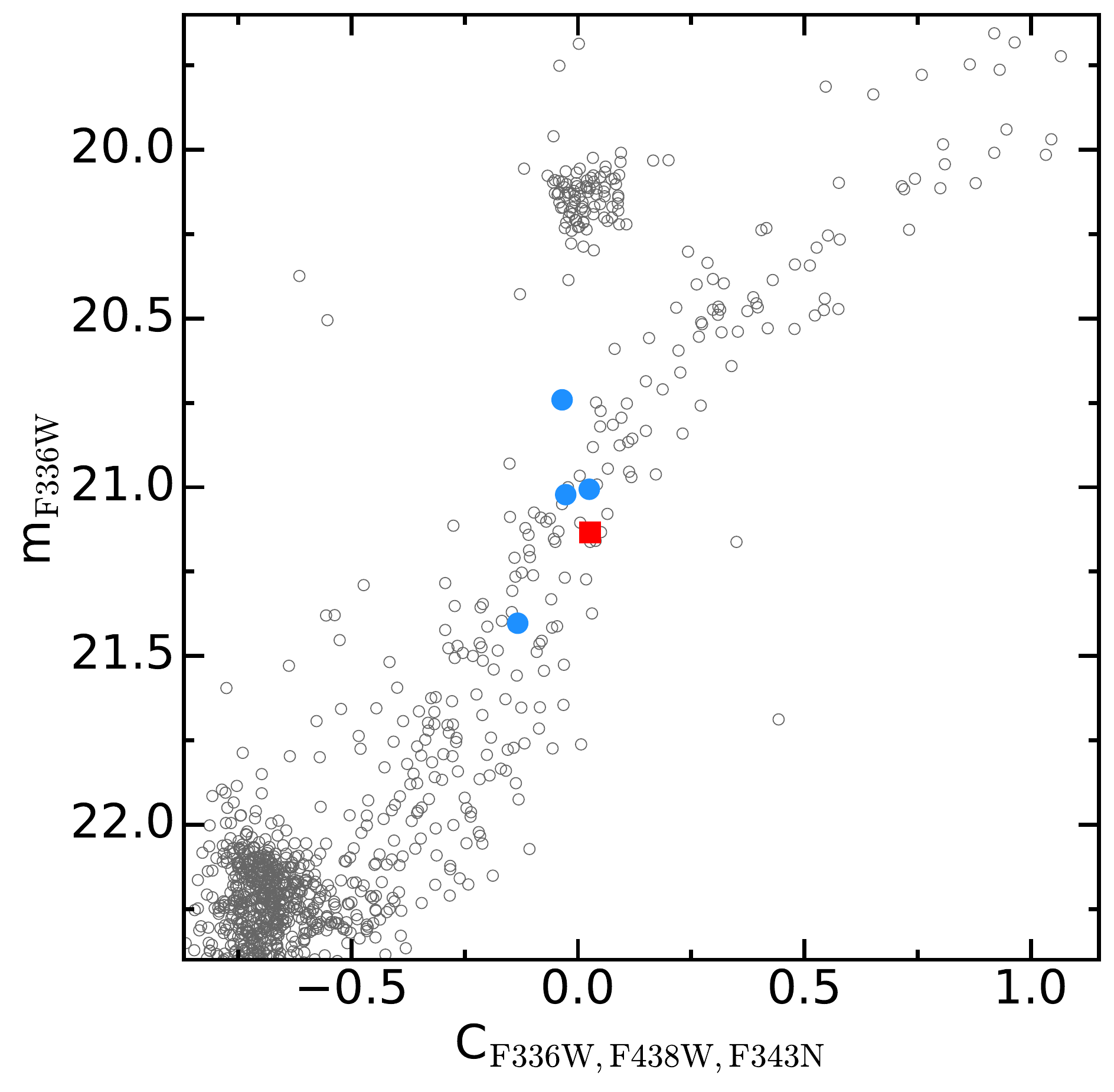}
  \caption{CMD of the RGB of Lindsay~1. The stars that we have in common with the sample from \citet{Hollyhead16} are marked as blue filled circles (primordial composition) and a red filled square (enriched). The star enriched in N lies on the red sequence whereas the stars with a primordial composition follow the blue sequence.}
   \label{fig:l1stars}
\end{figure}

Recently, \citet{Hollyhead16} spectroscopically studied the N and C content of 16 RGB stars that have been identified to be members of Lindsay~1. They found that six stars in their sample are significantly enhanced in N compared to the other stars, e.g. up to [N/Fe]$\sim$1.3~dex. The analysed stars are fainter than the bump in the luminosity function. Therefore internal mixing has a negligible impact in changing their C and N content, from which they conclude that the enhancement in N is evidence of a second population within Lindsay~1. We can now compare their spectroscopic results with our photometric data as a consistency check. For this, we cross-correlated the coordinates of the stars in their sample with the stellar positions in our photometric catalogue. We found that in total there are five cluster stars analyzed by \citet{Hollyhead16} that are within our field-of-view. One of them was determined to be enhanced in N, the other four have a primordial composition. We marked the stars in our CMD of Lindsay~1 (see Figure~\ref{fig:l1stars}). The red filled square is the star enriched in N, whereas the blue filled circles indicate primordial stars. We can see that the enriched star lies on the faint sequence whereas all stars with a primordial composition follow the brighter RGB. This was as expected, as the models predict that second population stars (stars enhanced in N) follow a redder RGB sequence. 

\section{Discussion and Conclusions}
\label{sec:conclusions}

We presented HST photometry for the three SMC intermediate-age clusters Lindsay~1, NGC~339 and NGC~416. Using the pseudo colour index $C_{F336W,F438W,F343N}$, we were able to detect a splitting in their RGBs, characteristic of the presence of multiple populations within the clusters. The results obtained in this paper provide an independent confirmation of the findings by \citet{Hollyhead16} who spectroscopically detected variations in N abundances in a sample of RGB stars in Lindsay~1. 

We determined the fraction of second population stars within the innermost regions of the clusters to be between $\sim$25\% and $\sim$50\% which is lower than the median value in Galactic GCs where the enriched stars typically outnumber the primordial population by a factor of 2 \citep[e.g.][]{BastianLardo15}. We note, however, that the values in \citet{BastianLardo15} are mainly from spectroscopic analyses and sample mostly the outskirts of the clusters. 
In NGC~121, a SMC star cluster as well, the enriched stars also make up only $\sim$30\% of the total stellar mass \citep[][Paper~I]{Dalessandro16}. It seems unlikely that this low fraction of second population stars is related to the mass of the cluster, as NGC~121 has a mass of $\sim 3.5 \times 10^5$~M$_{\sun}$ \citep{Mackey03,McLaughlin05}, comparable to typical masses of Milky Way GCs. As star clusters in the SMC seem to have a systematically lower fraction of second population stars than Galactic GCs, it might be that this fraction is dependent on the environment in which the cluster has formed. 
It is difficult to evaluate whether the enriched stellar fraction in SMC clusters correlates with the mass of the cluster, as the fractions of enriched stars reported here are uncertain.
\citet{Mucciarelli09} spectroscopically studied three GCs in the LMC and found a Na-O anti-correlation within these clusters. The number of stars analyzed in each cluster, however, is too small to determine the fraction of enriched stars in LMC GCs.

Lindsay~1, NGC~339 and NGC~416 are so far the youngest star clusters known to show star-to-star variations in light element abundances. These clusters are significantly younger than the ancient Galactic and extragalactic GCs showing multiple populations. We even detect a splitting in the RGB of the very extended cluster Lindsay~1 therefore it appears that even low-density clusters can be capable of forming multiple populations. We also note here, that the density of a cluster alone is not a sufficient criterion whether a cluster is capable in hosting multiple populations. In order to characterize the compactness of a cluster, \citet{Krause16} defined the compactness index $C_5$ as $(M_{\star}/10^5M_{\sun})/(r_h/pc)$, where $M_{\star}$ is the stellar mass of the cluster and $r_h$ its half-mass radius. Using the values for Lindsay~1 as given in \citet{Glatt11}, we find a $C_5$ index of 0.12. The Galactic open cluster NGC~6791 (Berkeley~46) has a smaller mass than Lindsay~1 but the same $C_5$ index as Lindsay~1 \citep{Krause16}. However, the open cluster has been shown not to host multiple populations \citep{Bragaglia14}.

The fact that we detect a splitting in the RGB in all three clusters suggests multiple populations are also a common characteristic of clusters with ages as young as 6~Gyr, and possibly even younger. This implies that the occurrence of this feature which is typical for GCs is not an early cosmological effect. Whatever process is causing the formation of multiple populations must have acted down to a redshift of, at most, z=0.65, where the conditions for cluster formation are expected to be similar to that observed in galaxies with a high cluster formation rate today (there is no significant change in the star formation rate within the SMC 6 - 8~Gyr ago, \citealt{Weisz13}). Hence, we might expect multiple populations to be present in even young clusters, as well. This would provide a direct link between young massive clusters forming today and old GCs, as predicted by several theories of cluster formation \citep[e.g.][]{Schweizer87, Ashman92,Kruijssen15}. However, studies by, e.g. \citet{Mucciarelli08,Mucciarelli11,Mucciarelli14} have not found evidence of light element abundance spreads in LMC star clusters with ages less than 3~Gyr, based on ground based spectroscopic data. In order to confirm at which cluster ages multiple populations are present or absent, larger samples of data, including space-based, are needed. In future works, we will extend our study to star clusters with ages younger than 6~Gyr for which we have already data in hand.

\section*{Acknowledgements}

We, in particular FN, NB, VKP and IP, gratefully acknowledge financial support for this project provided by NASA through grant HST-GO-14069 from the Space Telescope Science Institute, which is operated by the Association of Universities for Research in Astronomy, Inc., under NASA contract NAS526555.
NB gratefully acknowledges financial support from the Royal Society (University Research Fellowship) and the European Research Council (ERC-CoG-646928, Multi-Pop). DG gratefully acknowledges support from the Chilean BASAL Centro de Excelencia en Astrof\'{i}sica y Tecnolog\'{i}as Afines (CATA) grant PFB-06/2007. 
MJC gratefully acknowledges support from the Sonderforschungsbereich SFB 881 "The Milky Way System" (subproject A8) of the German Research Foundation (DFG).
We are very grateful to Jay Anderson and Andrea Bellini for sharing with us their ePSF software.
We thank the anonymous referee for useful comments and suggestions.




\bibliographystyle{mnras}
\bibliography{references} 







\bsp	
\label{lastpage}
\end{document}